\documentstyle[amsfonts,aps]{revtex}

\begin{document}
\title{Nonlocality, singularity, and elastic scattering in quantum fields}
\author{Hai-Jun Wang$^{*}$}
\address{Center for Theoretical Physics and School of Physics, Jilin University,
Changchun 130023, China}
\maketitle

\begin{abstract}
Using path integrals we express the quantum nonlocality of AB-effect type in
the form of singularity. The gauge-fixing term in path integrals induce the
AB effect in ordinary scattering processes. This means that all scattering
processes are accompanied by nonlocal effect. The formulae are then extended
to theory of fields that additionally include a scalar potential. It turns
out that the degree of freedom of nonlocality in quantum fields is just the
degree of the ghosts. Furthermore, renormalization method can be related to
this type of nonlocal effect.

*E-mail address: whj@mail.jlu.edu.cn
\end{abstract}

{\it Introduction. }The nonlocality in quantum mechanics has long been a hot
topic in the past decades, and up to date there has been no experiment
contradicting the nonlocality; It refers to the correlation between two
particles separated in space such as entanglement derived from the Bell
theory [1] and well confirmed in many experiments [2]. All these experiments
used massless photons as carriers of the states, and the nonlocality is of
the Bell type. The study of nonlocality has been also extended to a single
photon, which is in a superposition state of two space-separated states such
as $\mid a\rangle $ and $\mid b\rangle $, as in the situation when a photon
passes through a two-slit plane (diffraction). This type of nonlocality is
called Hardy type [3, 4]. The conventional method of studying this type of
states is to introduce two vacuum states $\mid 0\rangle _A$ and $\mid
0\rangle _B$ at two local regions for $A$ and $B$ [5-7]. Then the
correlation of states $\mid a\rangle $ and $\mid b\rangle $ is obvious when
one couples them to entangled state $\mid \Psi \rangle =\frac 1{\sqrt{2}}%
(\mid a\rangle \mid 0\rangle _A$ $+e^{i\phi }\mid b\rangle $ $\mid 0\rangle
_B)$. However, some authors argued that [8] for a single fermion, the above
method is not effective any longer because the massive fermion manifests its
nonlocality in a completely different way. Since its nonlocality can't be
transformed to explicit correlation as above, one can't measure it at
location $A$ and $B$ and then evaluate how it violates the Bell
inequalities. Through a tedious analysis, the authors arrived at the
conclusion that the only type of nonlocality for a fermion (except the
collapse) wave is of the AB-effect [9] type.

The AB effect for fermions has been well demonstrated by experiments [10,
11, 12]. It appears when a charged particle winds around a magnetic flux
completing closed {\it n }(integer) loops. To understand the effect in more
general circumstances [13], one may ask what will happen if a charged
particle is scattered by a very thin flux such as a spin moment. The answer
to this question provides us with some more useful observables of
nonlocality other than the familiar scattering results. The scattering
amplitudes and phase shifts have been studied quantitatively [14, 15], and
distinguishing feature is that the scattering result loses the axile
symmetry and dependence on the magnetic component appears. But, how does the
nonlocal effect take place {\bf simultaneously} with the scattering process
remains elusive as does not make sense to require the charged particle to
move around the flux for exactly {\it n} closed loops in a certain plane and
then come back to fulfill the scattering process involving other
interactions.

In this paper we will demonstrate that the nonlocal effect takes place when
and only when scattering occurs. First, we prove that the Aharonov-Bohm
effect can be reproduced if we introduce a singularity to the Feynman path
integral for (2+1)-dimensional quantum electrodynamics. Secondly, the
condition of (2+1)-dimension can be removed by considering the gauge
transformation and gauge-fixing condition. To this end, the nonlocality like
AB-effect can be described using the same Lagrangian with only an additional
singularity. In above expression, we have viewed the fermion that is with
magnetic moment as rest singularity, the other fermion is initial and final
particle that being scattered. This view makes us aware that any scattering
process is accompanied by the AB-like nonlocality. The resultant formulae
can be easily generalized to quantum field and thus to the non-Abelian
situation. It is showed that the degree of nonlocality is relevant to the
degree of freedom of ghost fields. The application to the renormalization
group is sketched.

In what follows, we employ the path integral method of Feynman to interpret
the propagating process of the wave function [16]. For instance, using the
kernel $K({\bf x}_2,t_2;{\bf x}_1,t_1)$ to describe how the wave function $%
\psi ({\bf x}_2,t_2)$ has evolved from all states $\psi ({\bf x}_1,t_1)$ at
the moment $t_1$: 
\begin{equation}
\psi ({\bf x}_2,t_2)=\int K({\bf x}_2,t_2;{\bf x}_1,t_1)\psi ({\bf x}%
_1,t_1)d^3{\bf x}_1
\end{equation}
and 
\begin{equation}
K({\bf x}_2,t_2;{\bf x}_1,t_1)=\int [{\bf d}q]e^{iS}
\end{equation}
where $[{\bf d}q]$ denotes all possible paths, $S=$ $\int%
\nolimits_{t_1}^{t_2}L(q,\dot q){\bf d}t$ is the action along a certain
path, and $L(q,\dot q)$ is the corresponding Lagrangian.

In calculation, the following two properties of the kernel are very useful.
In quantum mechanics level, the main contribution to the kernel comes from
the paths that nearly make the action $S$ in an classical extremum $S_{cl}$
[16] up to a normalizing factor, i.e. $K\sim e^{iS_{cl}/\hbar }$if the $L(q,%
\dot q)$ in $S$ is a quadratic form. And due to the kinetic energy $\frac 12%
m\,\dot q^2$, few situations violate the quadratic requirement. Although
this simplification originally appeared only as a mathematical technique, it
can also simplify our physical consideration evidently, which will be
validated later.

Another feature of kernel is that ''{\it Amplitudes for events occurring in
succession in time multiply}.'' [16]. For two such succeeding events $%
a\rightarrow c$ and $c\rightarrow b$, we have 
\begin{equation}
K(b,a)=\int_{x_c}K(b,c)K(c,a){\bf d}x_c
\end{equation}

(I) {\it Expressing AB effect in Feynman Integral with a singularity}: we do
the integral of kernel as usual for any Lagrangian, and then evaluate the
contribution from the singularity and its neighborhood. By virtue of Eq.
(3), we can perceive the removal of singularity as a succeeding event after
the just finished integral. This is equal to the case that a particle walks
backward along all the original paths near the singularity after the
particle has already completed all the possible paths. Henceforth we call
this sort of walking backward as a {\bf return }mechanism. So, in physical
sense, the redundant integral performed around the singularity is removed
not by subtracting but by multiplying another kernel. Although the events
are in successive order and the latter motion is along the same original
paths, the integral over the position space should be written out using
different variables for distinguishability. 
\begin{eqnarray}
K(b,a)_{total} &=&K_s(\text{singular region})K(b,a)  \nonumber \\
\ &=&\int [{\bf d}q^{\prime }]e^{i\int\nolimits_{t_b}^{t_c}L(q^{\prime },%
\dot q^{\prime }){\bf d}t}\int [{\bf d}q]e^{i\int\nolimits_{t_a}^{t_b}L(q,%
\dot q){\bf d}t}  \nonumber \\
\ &=&\int [{\bf d}q][{\bf d}q^{\prime }]e^{iS+iS^{\prime }}
\end{eqnarray}
the measure $[{\bf d}q]$ stands for all the paths and $[{\bf d}q^{\prime }]$
represents all the paths covering the neighborhood of the singularity, see
Graph 1. It will be clear later that the action $S$ is associated with
scattering calculation and $S^{\prime }$ associated with nonlocal effect.

In general, we are only concerned with the effect of singularity on the wave
function in Eq.(1). In such a case we only need to evaluate the first kernel 
$K_s=\int [{\bf d}q^{\prime }]e^{iS^{\prime }}$. Now suppose an ideal
dimensional (2+1) situation in which a low energy electron is scattered by a
neutron: a plane is predetermined and the magnetic moment of neutron is
vertical to the plane. A general Lagrangian has a simple form as $L=$ $\frac 
12m\,\dot q^2-e(\varphi -\stackrel{\rightharpoonup }{v}\cdot \stackrel{%
\rightharpoonup }{A})$, in which $\varphi $ and $\stackrel{\rightharpoonup }{%
A}$ are scalar and vector potential respectively. For neutron, we know $%
\varphi =0$ and make $\stackrel{\rightharpoonup }{A}=(x,y,0)$. Then it is
straightforward to carry out the integral $K_s\sim e^{iS_{cl}^{\prime }}$
using the above-mentioned first property. But the result is not what we have
expected for the singularity. In a plane with singularity, the contribution
to the kernel can't be approximated using only one classical extremum of $%
S_{cl}^{\prime }$. It is obvious that the paths encompassing a singularity
can't be topologically invariably shrunk to one classical path. This point
will be clearer with the consideration of gauge fixing condition{\it .}
Using the above mentioned $K\sim e^{iS_{cl}},$ we may get 
\begin{equation}
K_s=\int ([{\bf d}q^{\prime }]_{\text{left}}+[{\bf d}q^{\prime }]_{\text{%
right}})e^{iS_{cl}^{\prime }}\sim e^{iS_{cl}^{\text{left}}}+e^{iS_{cl}^{%
\text{right}}}\text{.}
\end{equation}
In the integral of action $S_{cl}^{\prime }=\int L(q,\dot q){\bf d}t=\int (%
\frac 12m\,\dot q^2+\stackrel{\rightharpoonup }{v}\cdot \stackrel{%
\rightharpoonup }{A}){\bf d}t$, the second term $\int \stackrel{%
\rightharpoonup }{v}\cdot \stackrel{\rightharpoonup }{A}{\bf d}t=\int 
\stackrel{\rightharpoonup }{A}\cdot {\bf d}\stackrel{\rightharpoonup }{r}$
can just afford the phase difference $K_s\sim e^{iS_{cl}^{\text{left}%
}}(1+e^{i/\hbar 
\textstyle \oint
\stackrel{\rightharpoonup }{A}\cdot {\bf d}\stackrel{\rightharpoonup }{r}})$
required by AB-effect. However, there yet left the first term to be treated
in consistency. By a similar process, we obtain 
\begin{equation}
\int \frac 12m\,\dot q^2{\bf d}t=\frac 12\int \stackrel{\rightharpoonup }{p}%
\cdot {\bf d}\stackrel{\rightharpoonup }{r}\text{,}
\end{equation}
where $\stackrel{\rightharpoonup }{p}$ is the momentum, and thus the phase
difference is 
\begin{equation}
\frac 12%
\textstyle \oint
\stackrel{\rightharpoonup }{p}\cdot {\bf d}\stackrel{\rightharpoonup }{r}=%
\frac 12\int (\nabla \times \stackrel{\rightharpoonup }{p})\cdot {\bf d}%
\stackrel{\rightharpoonup }{\sigma }
\end{equation}
(${\bf d}\stackrel{\rightharpoonup }{\sigma }$ is differential area
element), which actually vanishes when the momentum is constant. Obviously
when a momentum is split into two on one side and merge on another side of
the singularity, the direction of momenta along the two paths will change.
Using the method of difference, we can prove that $(\nabla \times \stackrel{%
\rightharpoonup }{p})$ is non-vanishing (Graph 2), so is the nontrivial
difference. This difference from $(\nabla \times \stackrel{\rightharpoonup }{%
p})$ is expected to be responsible for the phase shift due to diffraction,
which should be held even when the interaction $\stackrel{\rightharpoonup }{v%
}\cdot \stackrel{\rightharpoonup }{A}$ is absent. The interaction of scalar
potential $\varphi $ can be included (neutron replaced by proton) without
affecting the above result since it is path-independent.

(II) {\it Generalizing the expression of nonlocality}. In order to
generalize the expression, it is necessary to consider the gauge
transformation $\vec A\rightarrow \vec A^{\prime }=\vec A+\nabla \theta (x)$%
, and gauge fixing condition, e.g. coulomb gauge $\nabla \cdot \vec A=0$.
Concerning the gauge transformation, since the scalar function $\theta (x)$
can be arbitrarily determined, it can directly reduce the degree of freedom
of $\vec A$ by one. Therefore, the vector $\vec A$ can directly be chosen as 
$\vec A=(\vec A_x,\vec A_y,0)$, and thus in a plane. On the other hand, we
know that any two-particle scattering must occur in a plane which can't be
predetermined but actually exists. So as an effective choice (equivalent
way), make all the $\vec A$'s are parallel to the scattering plane. So, the
above assumption for AB effect that the scattering must happen in a plane
can actually be removed. Additionally, $\nabla \cdot \vec A=0$ suggests that
the vector $\vec A$ behaves like a tangent of magnetic lines without
divergence; and hence form closed loops. To this end, the condition of
AB-effect, in a plane and close loop, is automatically satisfied by
two-particle scattering with regard to the gauge transformation and gauge
fixing condition. These two constraints reduce the freedoms of vector $\vec A
$ from three to one, identical to the freedom along a loop. We note that the
freedom for gauge transformation is the very nonlocal degree of the fermion.

On an alternative viewpoint, since the choice of function $\theta (x)$ for
the gauge transformation $\vec A\rightarrow \vec A^{\prime }=\vec A+\nabla
\theta (x)$ is arbitrary, $\vec A$ can be replaced by $\nabla \theta (x)$ at
every point in a continuous manner. Thus the discussion of $\vec A$ also
applies to $\nabla \theta (x)$, i.e. the $\nabla \theta (x)$ is in the
scattering plane with $\nabla \cdot \nabla \theta (x)=0$. A change of $%
\nabla \theta (x)$ in $\vec A$ transform the wave function ${\em \psi }$ to $%
{\em \psi }e^{i\theta (x)}$. The movement will not be trivial if a
singularity exists for it produces observable interference stripe of the
AB-effect.

In QED the coulomb gauge is equivalent to the Lorentz gauge $\partial _\mu
A^\mu =0$, so the above argument {\it can be extended to include scalar
potential}. We do this at the cost of giving up the time arrow, and the
''plane'' in which scattering and diffraction occur is now a 3-dimensional
''super plane''. The condition $\nabla \cdot \nabla \theta (x)=0$ becomes $%
\partial _\mu \partial ^\mu \theta (x)=0$, accordingly. Similarly, the two
constrains $A_\mu \rightarrow A_\mu ^{\prime }=A_\mu +\partial _\mu \theta
(x)$ and $\partial ^\mu A_\mu =0$ will reduce the freedoms of $A_\mu $ from
four to two. In this way, the scattering considered here is automatically
extended to the general cases of a two-fermion scattering.

To extend the above argument to Field theory, let's write down the gauge
condition in terms of Faddeev and Popov form [17]. In QED, the form is

\begin{equation}
\triangle [A]\int [d\theta ]\delta (\partial _\mu (A^\mu )^\theta )=1,
\end{equation}
where $\triangle [A]$ is the Jacobian 
\begin{equation}
\frac{\delta (\partial _\mu (A^\mu )^\theta )}{\delta \theta }=\partial _\mu
\partial ^\mu \delta (x-y)=\Box \delta (x-y)\text{,}
\end{equation}
which is independent of $A^\mu $ in Abelian case, and hence gauge invariant.
Eq. (8) is demanded to be gauge invariant, so the gauge invariance of $\int
[d\theta ]\delta (\partial _\mu (A^\mu )^\theta )$ automatically restores
the result $\partial _\mu \partial ^\mu \theta (x)=0$. The gauge condition $%
\partial _\mu A^\mu =0$ can be controlled in experiments by specifying the
final outcome, but the resultant path of $\partial ^\mu \theta (x)$ is a
pure gauge property that can't be fixed by using local experiments, i.e. the
function $\theta (x)$ can't be determined locally. If considering the
integral for the Grassman variables $c_i$, $(\prod_{i=1}\int
dc_i^{*}dc_i)e^{-c_i^{*}B_{ij}c_j}=$det$B$ [18], $\triangle [A]$ in eq. (8)
can be expressed using degree [18] of freedom of the ghost as 
\begin{equation}
\int [{\cal D}\bar c][{\cal D}c]e^{i\int d^4x\,\bar c(x)\Box c(x)}
\end{equation}
in the case of QED the above integral only contains dynamical term that can
be absorbed into normalization constant. We reserve its explicit form here
to see clearly that the freedom of ghosts (${\it \bar c(x)}$ and ${\it c(x)}$%
) is just the nonlocal degree of freedom. And the degree of freedom of an
electron is 4, larger than the ghosts.

From amplitude $K_{total}$ it is straightforward to extend the above
discussion to field theory. Notice the form of Eq.(1) and the expansion
property of quantum mechanics ${\em \psi }=\sum c_i\varphi _i$, every
element of the S-matrix can be written as

\begin{equation}
\langle in\mid out\rangle \sim K_{total}=K_sK_{normal}\text{.}
\end{equation}
Now the integral measurement should be replaced by the fermion and boson
fields instead of configuration space. For QED, $K_{normal}$ has the form $%
\int [{\cal D}{\em \bar \psi }][{\cal D}{\em \psi }][{\cal D}{\em A}]e^{iS}$%
, where the Lagrangian in $S$ now becomes ${\frak L}=-1/4\,F_{\mu \nu
}F^{\mu \nu }+{\em \bar \psi }{\it (i}{\em \not D}{\it -m)}{\em \psi }$ with 
$F_{\mu \nu }=\partial _\mu A_\nu -\partial _\nu A_\mu $ and $D_\mu
=\partial _\mu -ieA_\mu $ . A gauge-invariant Lagrangian made by imposing
gauge fixing condition and adding the ghost [19, 20] has a general form 
\begin{equation}
{\frak L}=-1/4\,F_{\mu \nu }F^{\mu \nu }+\bar \psi {\it (i\not D-m){\em \psi 
}+}\frac 12\xi (\partial ^\mu A_\mu )^2+\bar c(x)(\partial ^\mu \partial
_\mu )c(x)
\end{equation}
A Lagrangian with ghost field is convenient to do BRST transformation to get
identities used in renormalization. Here we skip the details along this
context. Notice that the term $\bar c(x)(\partial ^\mu \partial _\mu )c(x)$
in Eq. (12) and the term $\partial ^\mu \partial _\mu \theta (x)=0$ in
Eq.(8) are both responsible for gauge invariance after the gauge fixing term
having been added [19]. Comparing the freedom of $\partial _\mu \theta (x)$
and ghost field $c(x)$, it is found that the ghost degree is very the
nonlocal degree. In QED, this degree of freedom is 2, one for scalar, one
for vector.

The above formulae can be further extended to the non-Abelian case (e.g.
QCD) by only changing $F_{\mu \nu }$ and $D_\mu $ to $F_{\mu \nu
}^a=\partial _\mu A_\nu ^a-\partial _\nu A_\mu ^a+gf^{abc}A_\mu ^bA_\nu ^c$
and $D_\mu =\partial _\mu -igA_\mu ^a\lambda ^a$, in which the $\lambda ^a$
is the generator of the gauge group and $f^{abc}$ is the corresponding
structure constant. The terms in Lagrangian of $K_s$ is the same as $%
K_{normal}$; however, the integral measurement should be independent.
Furthermore, at the neighborhood of singularity, the fermion wave function
and thus the fermion field should vanish. Here let's again concentrate on
the degree of freedom for nonlocality.

For non-Abelian field, Eq.(8) should be changed to 
\begin{equation}
\int [d\alpha ]\delta (G(A^\alpha ))\text{det}(\frac{\delta G(A^\alpha )}{%
\delta \alpha })=1\text{,}
\end{equation}
in which the Jacobian det$(\frac{\delta G(A^\alpha )}{\delta \alpha })$ is
related to $A^\alpha $ and can't be taken out of the integral as $\triangle
[A]$ in QED. For QCD, the gauge transformation is 
\begin{eqnarray}
A_\mu ^a &\rightarrow &(A^\alpha )_\mu ^a=A_\mu ^a+\frac 1g\partial _\mu
\alpha ^a+f^{abc}A_\mu ^b\alpha ^c  \nonumber \\
\ &=&A_\mu ^a+\frac 1gD_\mu \alpha ^a
\end{eqnarray}
from this expression we can get the analogous result of QED in Lorentz
gauge: 
\begin{equation}
\frac{\delta G(A^\alpha )}{\delta \alpha }=\frac 1g\partial ^\mu D_\mu \text{%
.}
\end{equation}
Similarly Eq.(10) takes the form: 
\begin{equation}
\text{det}(\frac{\delta G(A^\alpha )}{\delta \alpha })=\int [{\cal D}{\it 
\bar c}][{\cal D}{\it c}]e^{i\int d^4x\,\bar c(x)(-\partial ^\mu D_\mu )c(x)}
\end{equation}
Accordingly gauge-invariant Lagrangian becomes [18] 
\begin{eqnarray}
{\frak L} &=&-1/4\,F_{\mu \nu }F^{\mu \nu }+{\it \bar \psi (i\not D-m)\psi +}%
\frac 12\xi (\partial ^\mu A_\mu ^a)^2+\bar c^a(x)(-\partial ^\mu D_\mu
^{ac})c^c(x)  \nonumber \\
a,c &=&1,2,\cdots ,8
\end{eqnarray}
The fact that the degrees of the ghost fields are responsible for the
nonlocal degrees is directly extended from QED. In QCD, the degree of
freedom for ghost fields is 16. Here the discussion of nonlocality is more
complex for the ghost fields are coupled with the vector field, which is one
of the main features in non-Abelian field theory. The relationship between
the wilson loop and Berry phase [21] belongs to 2-dimensional out of
4-dimension case of the spatial degrees. And the degree of quarks here is $%
4\times 3=12$, smaller than $16$. It may be of this reason that quarks are
confined.

(III) {\it New Hamiltonian}. Let's start from the Eq.(4). Since the two
Lagrangians in actions $S$ and $S^{\prime }$ have the same forms, the same
integrand will certainly induce the same integral resultant function, in
which only boundary values of $q$ and $q^{\prime }$ are different. For
example, if resultant function is $S(q)$ and $S^{\prime }(q^{\prime
})=S(q^{\prime })$, then for the first action, $S_1=S(q)\mid
_{t_a}^{t_b}=S(q(t_b))-S(q(t_a))$, and for the second, $S_2=S(q^{\prime
})\mid _{t_b}^{t_c}=S(q^{\prime }(t_c))-S(q^{\prime }(t_b))$. To finish the
integral of [$dq$] and [$dq^{\prime }$], let's divide the time interval into
many parts as, $[t_a,t_b]\sim [t_1,t_2,\cdots ,t_l,\cdots ,t_m,\cdots ,t_n]$%
. Then the integral measurement [$dq$] changes with the time division to $%
[dq]=dq_2dq_3\cdots dq_l\cdots dq_m\cdots dq_{n-1}$. For the division of $%
[t_b,t_c]$, we determine the intervals of time so coincident with the time
interval between the $[t_l,t_m]$ as to reflect the {\bf return} mechanism:
If $[t_b,t_c]\sim [t_1^{\prime },t_2^{\prime },\cdots ,t_n^{\prime }]$, then
make $q^{\prime }(t_c)=q(t_l)$, $q^{\prime }(t_{n-1}^{\prime })=q(t_{l+1})$, 
$q^{\prime }(t_{n-2}^{\prime })=q(t_{l+2})$, $\cdots $, $q^{\prime
}(t_b)=q(t_m)$, which means $[dq^{\prime }]=dq_{m-1}dq_{m-2}\cdots dq_{l+1}$%
. And thus applying the second property of Feynman integral and multiplying
every integral results together yield 
\begin{eqnarray}
K(b,a)_{total} &=&K_s(\text{singular region})K_{normal}  \nonumber \\
\ &=&\int [{\bf d}q][{\bf d}q^{\prime }]e^{iS+iS^{\prime }}  \nonumber \\
\ &=&\int dq_2dq_3\cdots dq_l\cdots dq_m\cdots
dq_{n-1}e^{i(S_{12}+S_{23}+\cdots +S_{n-1,n})}\times  \nonumber \\
&&\ \ \int dq_{m-1}dq_{m-2}\cdots
dq_{l+1}e^{i(S_{m-1,m-2}+S_{m-2,m-3}+\cdots +S_{l+2,l+1})}
\end{eqnarray}
where $S_{i-1,i}$ represents the integral for action between $[t_{i-1},t_i]$%
. It should also be noted that, according to the {\bf return} mechanism and
our design of division, $\int dq_{m-2}e^{i(S_{m-1,m-2}+S_{m-2,m-3})}$[for $%
K_s$]$=(\int dq_{m-2}e^{i(S_{m-3,m-2}+S_{m-2,m-1})}$[for $K_{normal}$]$%
)^{-1} $. Equally, if the integrand for $dq_l\cdots dq_m$ is $\int
dq_{l+1}\cdots dq_{m-1}\delta (\Omega -\Omega _0)e^{i(S-S^{\prime })}=1$($%
\Omega _0$ is a singularity region), then there will be no paths between the
region $[q_{l+1}\cdots dq_{m-1}]$. In other words, the paths that contribute
nothing to the Kernel will only add a factor $1$ to the Kernel. So $\int
dq_{l+1}\cdots dq_{m-1}\delta (\Omega -\Omega _0)e^{i0}=1$or $constant$ and
is absorbed in normalized factor. For this reason the integral of Eq.(18)
has an equivalent expression, 
\begin{equation}
\int [{\bf d}q][{\bf d}q^{\prime }]e^{iS+iS^{\prime }}=\int [{\bf d}%
q]e^{i(S-S_d)}=\int [{\bf d}q]e^{i\int ({\frak L}-{\frak L}^{\prime })dt}
\end{equation}
Now the integral measurement is over all the region of space, and $S_d$ is
the part which is nontrivial only in singular region. The same analysis
applies equal to quantum fields: changing the integral measurement and
extending the variables in ${\frak L}$ and ${\frak L}^{\prime }$ directly to
fields quantities, and make fermion fields vanish at the singularity point
correspondingly.

Eq.(11) suggests that future calculations may consist of two steps. First,
we calculate the scattering amplitude $K_{normal}$ as usually done; then, we
consider the $K_s$ to include nonlocal effect. To calculate the correction
of higher order than the tree level, then, the amplitude $K_{normal}$ and $%
K_s$ have to be considered together. For the case of ultraviolet divergence
in QED and the infrared behavior in QCD, the ''{\bf return}'' region for $%
K_s $ may just be the forbidden region for $K_{normal}$, whose contribution
should be cut off for its non-physical meaning. In this sense, we make the
Lagrangian in $S_d$ as the counter terms for that in action $S$ in $%
K_{normal}$, and the Lagrangian has the same form as that in $S$ with only
some renormalization constants multiplied to the corresponding terms. The
new Lagrangian means a new Hamiltonian, which contributes only to nonlocal
phase as in Berry phase, and doesn't contribute to the transitional
amptitude. In this sense, the renormalization and the nonlocality have been
unified to an expression.

Here we will not reiterate the lengthy procedure of renormalization
calculations. It is easy to recognized that the cutoff for
renormalizationare determined by the nonlocal region which we are really
concerned. Knowing the geometry of this region can enable us to distinguish
what is physical from what is non-physical, and justify the effective region
that renormaliztion group can be properly used. Using the relation $\lambda
\sim \frac 1p$ can only get a rough estimation in purterbative situation.
Further investigations on this issue are under way.

In above argument we have employed the case that electrons(or protons, as in
Ref [13]) are scattered by neutron, but with the generalized formulae, there
is no constraints for the involved particles. The argument can be readily
extended to any scattering of two particles that carry both flux and
charge[14, {\it commence (i)}].

In summary, in this paper we argue that any scattering process must be
accompanied by the nonlocal effect which may induce resultant observables.
And it is also achieved that the degree of freedom of nonlocality is
equivalent to the degree of the ghost field, and also the gauge degree of
freedom. The understanding of the renormalization along this context is also
sketched. Though having been unified in the same path integral frame, the
nonlocality (diffraction or AB-effect) is really different from the
scattering. The former is sensitive to measurement and thus unobservable
locally, but the latter can be detected anyway. Now the dissertation can be
unified using the gauge transformation even in higher-dimension Unitary
space.


\section{References}

\begin{references}
\bibitem{}  J. S. Bell, Physics (N.Y.) 1, 195 (1964).

\bibitem{}  A. Aspect, P.Grangier, G. Roger, Phys.Rev.Lett 47, 460(1981); 
{\it ibid.} 49, 91(1982); P.G. Kwiat, K. Mattle, H. Weinfurter, A.
Zeilinger, {\it ibid.} 75, 4337(1995);W. Tittel, J. Brendel, H. Zbinden,
N.Gisin, {\it ibid.} 81, 3563(1998); G. Weihs, T. Jennewein, C.Simon, H.
Weinfurter, A. Zeilinger, {\it ibid.} 81, 5039(1998); A. Aspect, Nature
(London) 398, 189(1999).

\bibitem{}  S. M. Tan, D. F. Walls, and M. J. Collett, Phys. Rev. Lett.
66.252(1991)

\bibitem{}  W. T. M. Irvine, J. F. Hodelin, C. Simon, and D. Bouwmeester,
Phys. Rev. Lett. 95, 030401 (2005);

\bibitem{}  G. Bj\"ork and P. Jonsson, Phys. Rev. A 64, 042106 (2001) and
references therein.

\bibitem{}  H. W. Lee and J. Kim, Phys. Rev. A 63, 012305 (2001) and
references therein.

\bibitem{}  M. A. Can {\it et al,} J. Opt. B: Quantum Semiclass. Opt. {\bf 7,%
} L1-L3 (2005);

\bibitem{}  Y. Aharonov and L. Vaidman, Phys. Rev. A {\bf 61}, 052108 (2000);

\bibitem{}  Y. Aharonov and D. Bohm, Phys. Rev. {\bf 115}, 485-491 (1959);

\bibitem{}  R. G. Chambers, Phys. Rev. Lett. {\bf 5}, 3(1960);

\bibitem{}  A. Tonomura {\it et al, }Phys. Rev. Lett. 56, 792(1986);

\bibitem{}  B. E. Allman {\it et al, }Phys. Rev. A 48, 1799 (1993) and
references therein;

\bibitem{}  T. T. Wu and C. N. Yang, Phys. Rev. D 12, 3845(1975), refer to
the part VII, discussion on the scattering of neutron and proton{\it ;}

\bibitem{}  F. Wilczek and Y. S. Wu, Phys. Rev. Lett. 65, 13(1990);

\bibitem{}  D. H. Lin, Phys. Rev. A 69, 052711 (2004);

\bibitem{}  R. P. Feynman and A. R. Hibbs, {\it Quantum Mechanics and Path
Integrals}, McGraw-Hill, Inc., 1965: pp. 29 and pp.60 and a following
citation, pp.37.

\bibitem{}  L. D. Faddeev and V. N. Popov, Phys. Lett. 25B, 29(1967);

\bibitem{}  M. E. Peskin and D.V.Schroeder, {\it Quantum Field Theory},
Addison-Wesley Publishing Company, 1995:pp300 and pp513.

\bibitem{}  M. Lavelle and D. McMullan, Phys. Rev. Lett. 71, 3758 (1993);

\bibitem{}  J. C. Su and H. J. Wang, Phys. Rev. C 70, 044003 (2004).

\bibitem{}  Y. Chen, B. He, J. M. Wu, Mod. Phys. Lett. A15, 1127(2000); F.
V. Gubarev, V. I. Zakharov, Int. J. Mod. Phys. A17, 157 (2002), and
references therein.
\end{references}
\end{document}